\documentclass{article}
\usepackage{ijcai21}

\usepackage{tikz}
\usetikzlibrary{positioning,shapes}
\usepackage{pgflibraryshapes}
\usepackage{times}
\usepackage{soul}
\usepackage{url}
\usepackage[hidelinks]{hyperref}
\usepackage[utf8]{inputenc}
\usepackage[small]{caption}
\usepackage{graphicx}
\usepackage{amsmath}
\usepackage{amsthm}
\usepackage{booktabs}
\usepackage{algorithm}
\newcommand{\ie}{\textit{i}.\textit{e}., }

\usepackage{amsmath,amsthm,amssymb}
\usepackage{graphicx}
\usepackage{subfigure}
\usepackage{booktabs} 
\usepackage{algpseudocode}
\usepackage{algorithm}
\usepackage[utf8]{inputenc}
\usepackage[english]{babel}

\usepackage{hyperref}
\usepackage{mathtools}

\newtheorem{assumption}{Assumption}

\title{Human-AI Collaboration with Bandit Feedback}

\author{
Ruijiang Gao$^1$
\and
Maytal Saar-Tsechansky$^1$\and
Maria De-Arteaga$^1$\and
Ligong Han$^2$\and\\
Min Kyung Lee$^1$ \And 
Matthew Lease$^1$
\affiliations
$^1$University of Texas at Austin\\
$^2$Rutgers University
\emails
\{ruijiang, ml\}@utexas.edu,
\{maytal.saar-tsechansky, dearteaga\}@mccombs.utexas.edu,
lh599@scarletmail.rutgers.edu,
minkyung.lee@austin.utexas.edu 
}

\begin{document}

\maketitle

\begin{abstract}
    Human-machine complementarity is important when neither the algorithm nor the human yield dominant performance across all instances in a given domain. Most research on algorithmic decision-making solely centers on the algorithm's performance, while recent work that explores human-machine collaboration has framed the decision-making problems as classification tasks. In this paper, we first propose and then develop a solution for a novel human-machine collaboration problem in a bandit feedback setting. Our solution aims to exploit the human-machine complementarity to maximize decision rewards. We then extend our approach to settings with multiple human decision makers. We demonstrate the effectiveness of our proposed methods using both synthetic and real human responses, and find that our methods outperform both the algorithm and the human when they each make decisions on their own. We also show how personalized routing in the presence of multiple human decision-makers can further improve the human-machine team performance. 
\end{abstract}

\section{Introduction}
Supervised learning and algorithmic decision-making have shown promising results when solving some tasks traditionally performed by humans~\cite{chen2018rise,swaminathan2015counterfactual}. Yet, most of these works focus on improving algorithm performance, and do not consider optimizing human-machine team performance~\cite{bansal2020optimizing}. Human-machine complementarity is particularly promising when neither the state-of-the-art algorithm nor the human yield dominant performance across all instances in a given domain; there are thus opportunities in such settings to develop methods for human-machine collaboration that exploit human-machine complementarity~\cite{wilder2020learning,madras2018predict,wang2020augmented}.

In this paper, we first propose and then develop a solution for a novel human-machine team complementarity under batch learning from bandit feedback (BLBF)~\cite{swaminathan2015counterfactual}. We consider decision-making tasks involving judgment of alternative course of actions to select the choices leading to the best outcomes, and where available historical data reflect a human decision-maker's past choices and the corresponding outcomes. To our knowledge, this is the first paper to propose and develop a human-algorithm hybrid solution for this problem.

Recent work on human-machine teams has considered the task of learning to defer~\cite{madras2018predict}, also framed as learning to complement humans~\cite{wilder2020learning}. We study this problem in the context of bandit feedback, where rewards and outcomes depend on the action taken. This requires estimating alternative course of actions and selecting the actions leading to the highest expected reward. Examples of such tasks include designing the most effective personalized treatment for a patient ~\cite{xu2016bayesian} or determining prices that maximize profit in revenue management~\cite{bertsimas2016power}.
Furthermore, because in many settings algorithms are meant to assist and improve decisions that humans have been responsible for in the past, learning is based on historical decisions made by human decision makers, with bandit feedback. Thus, differently from the typical setup assumed in supervised learning, outcomes are only observed for actions taken in the past; for example, rearrest can only be observed if a prisoner is released~\cite{ensign2018decision}, and evidence of child maltreatment may only be found if an investigation is opened~\cite{de2018learning}. We propose a human-machine hybrid that can be trained on historical decisions with bandit feedback to exploit human-machine complementarity in  decision performance, with the goal of producing decisions of higher reward. We refer to this problem as Human-AI BLBF (\textsc{hai-blbf}).


Our goal of developing human-machine collaboration for this decision task is different in important ways from the traditional task of BLBF. Work on learning from bandit data typically assumes that historical decisions were made either by humans or by an existing algorithm, and that the policy learned will be the sole decision maker for future instances. In \textsc{hai-blbf}, historical decisions are made exclusively by humans, while future decisions will be made by the proposed human-AI collaboration system. 
Figure~\ref{fig:dag} outlines the human-machine collaboration framework we consider to address the \textsc{hai-blbf} problem, where each decision instance is routed to either the algorithm or the human decision maker, so as to recommend the best action. Such system can be advantageous when neither the human nor the algorithm are superior decision makers over the entire domain. To optimally exploit the human-machine complementarity: (a) a predictive algorithm must specialize in instances where it can indeed achieve superior decisions to those made by the human, while allowing the human to handle instances for which the human is more likely to achieve better performance; (b) a routing algorithm must correctly route decisions to the entity that is most likely to yield the best outcome. In this paper, we make the following contributions:

\tikzset{node/.style={rectangle,fill=gray!10,draw,minimum size=0.8cm,inner sep=0.05cm} }
\tikzset{node/.style={rectangle,fill=gray!10,draw,minimum size=0.8cm,inner sep=0.05cm} }
\tikzset{arc/.style = {->,> = latex, , } }
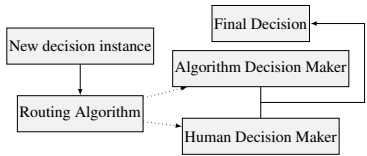
\begin{figure}
\centering
    \scalebox{0.6}{
    \begin{tikzpicture}[auto,node distance = 0.1 cm, scale = 1] 
    \node[node] at(-3,0.5) (S) { New decision instance };
    \node[node] at(-3,-1) (R) { Routing Algorithm };
    \node[node] at(1,0) (H) { Algorithm Decision Maker};
    \node[node] at(1,-1.5) (V) {Human Decision Maker};
    \node[node] at(1, 1) (Y) { Final Decision };
    \draw[arc] (S) to node{ } (R); 
    \draw[] (H) to node{ } (V); 
    \draw[dotted][arc] (R) to node{ } (H); 
    \draw[dotted][arc] (R) to node{ } (V); 
    \draw
    (1,-0.75) coordinate (A)
    (3.3,-0.75) coordinate (B)
    (3.3,1) coordinate (C)
    (2.1,1) coordinate (D)
    ;
    \draw[arc] (A) to (B) to (C) to (D);

     
    
    \end{tikzpicture}} 
    \caption{Decision Making Process for Human-AI Collaboration 
    }
    \label{fig:dag}
\end{figure}


\begin{itemize}
    \item We formulate a human-machine decision collaboration problem for bandit feedback, \textsc{hai-blbf}, where historical data reflect human decisions and corresponding outcomes, with the goal of exploiting the human-machine complementarity to achieve maximum total reward.  
    \item We propose a new counterfactual policy optimization objective to train a decision-making algorithm on observational data with the goal of complementing the human. At testing time, a routing algorithm assigns each decision instance to either a human or an algorithm decision-maker. The algorithm is jointly trained to best exploit the human's decision-making abilities, while complementing the human by optimizing the algorithmic decision-maker to perform well on instances that are harder for humans to achieve optimal choices. 
    \item We also propose a personalized variant for settings with multiple human decisions-makers, so as to also exploit humans' differential decision-making abilities.  
    \item We empirically demonstrate the performance of the proposed solutions for \textsc{hai-blbf} on multi-label datasets converted to reflect decision and outcomes, and using both synthetic and real human labels. 
    \item We investigate the limitations of our proposed hybrid team through ablation studies on model capacities, finding that a stronger model capacity is likely to decrease the reward improvement achieved by the hybrid system relative to what can be achieved by the model alone.
\end{itemize}

\section{Related Work}

\noindent \textbf{Human-AI Collaboration}
Recent research on human-AI collaboration shows that algorithms can outperform or help humans achieve better classification accuracy ~\cite{madras2018predict,wilder2020learning}. Empirical work considering humans-in-the-loop has studied how algorithmic risk assessment tools influence judges' decisions~\cite{stevenson2019algorithmic}, 
the impact of algorithmic risk assessment on decision-making criteria~\cite{green2020algorithmic}, and workers' ability to dismiss erroneous algorithmic recommendations in child maltreatment hotline screening~\cite{de2020case}.~\cite{bansal2020optimizing} study the problem of human-AI interaction and find that the best classifier is not always the one that leads to the best human decisions when the classifier is used as decision support. In this regard, we note that our proposed system does not involve human-machine interaction, and instead decisions are made by either the human or the algorithm.
The closest problems to \textsc{hai-bfbl} were considered in ~\cite{madras2018predict,wilder2020learning,raghu2019algorithmic,de2020regression,wang2020augmented}, where the problem of routing instances to an algorithm or a human was investigated. To allocate an instance to either the human or the machine, ~\cite{madras2018predict,raghu2019algorithmic,wilder2020learning} considered the prediction uncertainty of the algorithm and the human for that instance. ~\cite{de2020regression} studied human-machine collaboration for a regression task, and ~\cite{wang2020augmented} consider jointly augmenting humans' accuracy and fairness. The core difference between these works and ours is that they considered supervised learning tasks, whereas we focus on the problem of selecting an optimal policy given bandit feedback, \ie a  counterfactual policy optimization problem. 

\noindent \textbf{Counterfactual Policy Optimization}
Counterfactual policy optimization has received a lot of attention in the machine learning community in recent years~\cite{swaminathan2015counterfactual,joachims2018deep}. 
Counterfactual Risk Minimization (CRM), also known as off-policy learning or batch learning from bandit feedback, typically utilizes inverse propensity weighting (IPW)~\cite{rosenbaum1987model} to account for the bias in the actions reflected in the data. ~\cite{swaminathan2015counterfactual} introduces the CRM principle with a variance regularization term derived from an empirical Bernstein bound for finite samples. In order to reduce the variance of the IPW estimator, ~\cite{swaminathan2015self} proposes a self-normalized estimator, while BanditNet~\cite{joachims2018deep} proposes an estimator that is easy to optimize using stochatic gradient descent for deep nets. In this paper, we extend the traditional CRM problem and propose a method to incorporate human-algorithm decision complementarity in the optimization; we show that human-machine collaboration can further improve both the human and algorithm's performance, and that a joint optimization and personalized routing may lead to a further improvement. Our setup is also close to ensemble bandits~\cite{pacchiano2020model} to identify the optimal bandit algorithm in an online fashion and our goal is to learn the best hybrid system offline. 

\noindent \textbf{Personalizing for Diverse Worker Skills} Many supervised learning tasks require human labeling, which is often imperfect ~\cite{huang2017cost,yan2011active}. Recent research has begun to study label crowdsourcing with multiple noisy labelers and considered a variety of settings and goals ~\cite{yan2011active,huang2017cost}. For example, ~\cite{yan2011active} proposed a probabilistic framework to infer the label accuracies of workers and choose the most accurate worker for annotation; \cite{huang2017cost} allocated diverse workers with different costs and labeling accuracies to cost-effectively acquire labels for classifier induction. All these works, however, consider the context of supervised classification tasks, and human input is limited to labeling historical data for model training, rather than performing decision-making at test time. 
Decision-theoretic active learning ~\cite{nguyen2015combining} also considers human-machine hybrid for crowd-labeling for classification. 
Our work focuses on training models from bandit feedback, and considers that humans can be asked to make decisions at testing/deployment time, and thereby are integral to the human-machine decision-making team. 

\section{Problem Statement}

We use $\mathcal{X}$ to represent an abstract space and $\mathbb{P}(x)$ is a probability distribution on $\mathcal{X}$. Each sample $x = x_1, \ldots, x_n \in \mathcal{X}^n$ is drawn independently from $\mathbb{P}(x)$. $\mathcal{A}$ is the discrete action space from which a central agent can select an action for each sample, immediately after which a reward $r\in[0,1]$ is revealed to the agent. We assume rewards are revealed immediately after action is taken and leave for future work settings where rewards are revealed with a delay. For example, in precision medicine, $\mathcal{X}$ may represent a patient cohort, $\mathcal{A}$ refers to all possible treatments for a disease, and $r$ may refer to the treatment effect. 
Similarly, for personalized pricing, $\mathcal{X},\mathcal{A},r$ can represent customers, personalized discount rate, and revenue generated, respectively.
In this paper, we also assume a human decision-maker incurs a cost $C(x_i)$ when making a decision, reflecting the time, resources or compensations that might be required for a human to solve the task.


\noindent \textbf{Counterfactual Risk Minimization (CRM)}
Here we introduce the CRM objective. We focus on an example of precision medicine to illustrate our problem. We use $x \in \mathcal{X}^n \sim \mathbb{P}(x)$ to denote a patient. Let $\mathcal{A}$ represent discrete treatment options that a central agent can offer to patients. After offering treatment $a\in \mathcal{A}$, the agent observes the response from the patient $r\sim \mathbb{P}(r|x,a)$, \ie the treatment effect. Observational data is assumed to be generated by a randomized policy $\pi_0(\cdot|x)$, which may be generated by some human decision makers $h_i \in \mathcal{H}$. The goal is to maximize the expected reward under a learned policy $\pi_\theta(a|x)$: 
\begin{align}
\mathbb{E}_{x\sim\mathbb{P}(x), a\sim\pi_\theta(a|x)}r(x,a)    
\end{align}
We can optimize $\theta$ over the observational data using importance sampling~\cite{swaminathan2015counterfactual,rosenbaum1987model}, since

\begin{align}
    \mathbb{E}_{x\sim\mathbb{P}(x), a\sim\pi_\theta(a|x)}r(x,a)= \mathbb{E}_{x\sim\mathbb{P}(x), a\sim\pi_0(a|x)}\frac{\pi_\theta(a|x)}{\pi_0(a|x)}r(x,a)
\end{align}

\noindent thus we can optimize the objective on observed data by 
    \begin{align}
    \label{eqn:ips}
        \max_\theta \frac{1}{N}\sum_i r_i \frac{\pi_\theta(a_i|x_i)}{\pi_0(a_i|x_i)}
    \end{align}

If $\pi_0(a|x)$ is not known a priori, it is estimated by an additional classifier $\hat{\pi}_0(a|x)$ trained on historical data. With observational data, the individualized treatment effect is not always identifiable. We use Rubin's potential outcome framework and assume consistency and strong ignorability which is a sufficient condition for identifying ITE~\cite{pearl2017detecting}. Here we formally present the ignorability assumption~\cite{rubin2005causal}:
\begin{assumption}[Ignorability]
Let $\mathcal{A}$ be action set, $x$ is context (feature), $r(a)|x$ is observed reward for action $a\in\mathcal{A}$ given context $x$, $r(a) \perp\!\!\!\perp a | x, \forall a \in \mathcal{A}$.
\end{assumption}
In other words, we assume there are no unobserved confounders. Establishing if this condition is met generally cannot be achieved purely based on data and requires some domain knowledge. Next we discuss the proposed collaboration objective for the human-in-the-loop system, illustrated in Figure~\ref{fig:dag}. 

\noindent \textbf{Collaboration Objective}
Humans are imperfect decision-makers, while models are often of limited capacity, meaning that improving performance in some regions of the feature space sacrifices performance in others. This calls for the question: Can a human-AI collaboration exploit possible complementarities, producing better decisions and higher rewards by allocating tasks to either human decision-makers or the algorithm? Assuming we have learnt an algorithm $\pi_\theta(\cdot|x)$ from the observational data using equation~\ref{eqn:ips}, we can use an additional routing algorithm $d_\phi(h|x)$ to decide whether an instance $x$ should be assessed by a human decision maker or by the algorithm, as shown in Figure~\ref{fig:dag}. Since we have an additional option to choose whether a human or an algorithm should assess each instance, the hybrid system's reward relies on $d_\phi(h|x)$. Thus, the task is to learn a policy $d_\phi(h|x)$ by maximizing the objective 

\begin{align}
\label{eqn:2s}
\max_{\phi}\sum_{i=1}^N d_\phi(h|x_i)(r_i - C(x_i))  + \frac{  (1-d_\phi(h|x))\pi_\theta(a_i|x_i)}{\hat{\pi}_{0}(a_i|x_i)}r_i   
\end{align}

 \noindent where $C(x)$ is the additional cost a human decision-maker will incur, such as time and resources required by experts. Accounting for this cost may be especially useful in contexts where human expertise are limited (e.g. medical specialists during a pandemic), but the cost can also be set to zero in contexts where the best decision is desired, irrespective of the labor cost. For each instance $x_i$, we have a probability $d_\phi(h|x_i)$ to select a human decision maker or $1-d_\phi(h|x_i)$ to use the algorithm's decision, the objective thus is a weighted average of human and algorithm reward. We refer to the objective in Equation~\ref{eqn:2s} as Two-Stage Collaboration (TS) since the policy and the routing model are trained sequentially. 
 
\begin{figure}[h]
    \centering
    \includegraphics[width=0.6\linewidth]{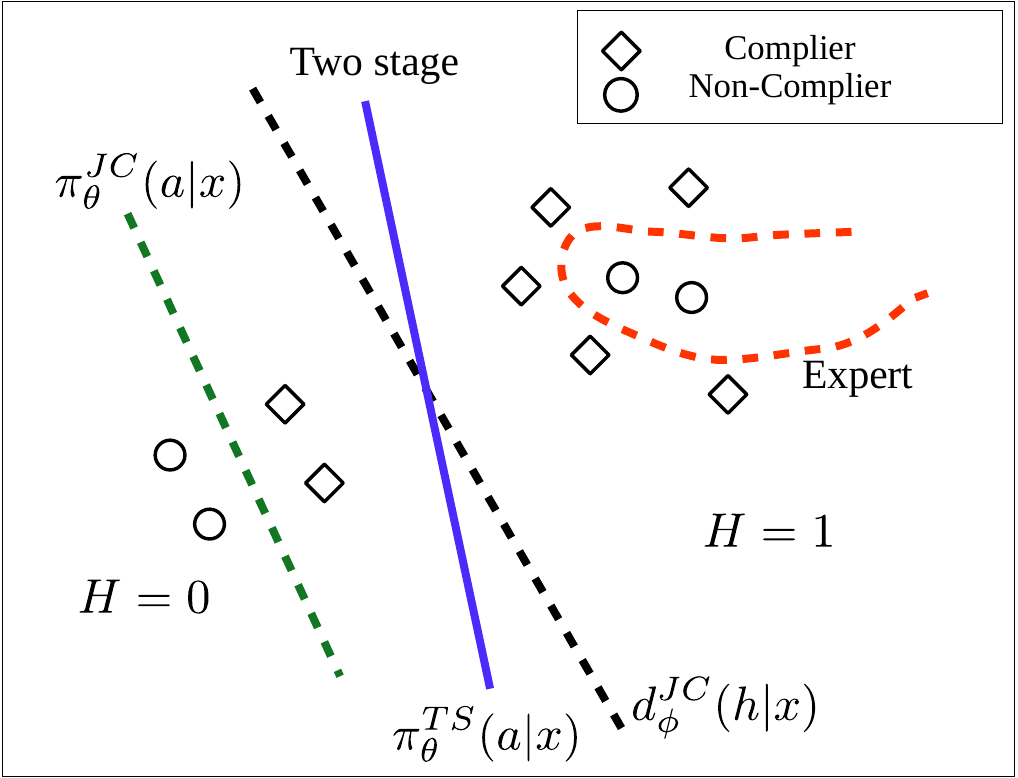}
    \caption{Joint Learning versus Two Stage Procedure: Dashed line represents the solution from joint learning (Green: Algorithm, Red: Expert, Black: Router) and the blue solid line represents the fixed algorithm solution from the two stage procedure.}
    \label{fig:jl_ex}
\end{figure}

 We are also interested in whether a joint optimization of both models could further improve the human-machine collaboration performance by optimizing Equation~\ref{eqn:2s} with respect to $\theta$ and $\phi$ jointly.
Thus, instead of fixing $\theta$ first, and then optimizing for $\phi$, we jointly optimize $d_\phi$ and $\pi_\theta$. 
Here we use a simple example, shown in Figure~\ref{fig:jl_ex}, to illustrate how the joint modeling procedure may outperform the two stage baseline. A similar example for classification is proposed in ~\cite{mozannar2020consistent}, and here we show its connection to counterfactual policy optimization.  Consider a setting where compliance of the treatment is not perfect~\cite{angrist1996identification}, and the treatment has a positive effect for all instances. If the treatment and outcome are both binary, the policy optimization problem simply reduces to binary classification, since it is sufficient to identify which candidates will comply with the treatment. Assume the feature space is two dimensional, the model class is chosen to be hyperplanes, compliers and non-compliers to the treatment are distributed as shown in Figure~\ref{fig:jl_ex}, and experts use a non-linear decision boundary to classify candidates. One may also assume experts have additional side information~\cite{mozannar2020consistent} that is not accessible to the algorithm. However, we do not make this assumption in this paper since it will violate our assumption of ignorability. Dashed lines represents the solution from the joint learning approach (Green: Algorithm, Red: Expert, Black: Router), which classifies all candidates correctly. However, if we use a two stage procedure, and given the model class has limited capacity, we can get an algorithm similar to the blue solid line, and fix it; consequently, one possible solution for the routing model is a hyperplane to the left of all points, assigning all tasks to the (more costly) human, and yielding sub-optimal performance.  

\noindent \textbf{Personalization Objective}
In the collaboration objective we discussed above, we assume experts are homogeneous.
However, in practice, different experts often have different expertise. For instance, one general physician may be more effective in treating older patients with comorbidities, while another may have greater expertise in treating younger adults. Therefore a personalized routing model might further improve the human-machine team performance. 

Instead of randomly assigning an expert to assess an instance, the routing algorithm can decide whether to refer the instance to an algorithm or to a human, and more specifically, which human decision-maker should be selected. 

One challenge of this problem is that we typically do not have access to human decision makers' policy $\pi_0(p|x,h_i)$. Thus, we model decision makers' policy with a supervised learning model conditioned on expert identity $h_i$ and pre-train it on observational data to approximate it. An alternative would be to train a separate supervised learning model for each human decision maker. Then we can train a \textsc{hai-blbf} algorithm by 


\begin{align}
    \label{eqn:jointp}
    \max_{\theta, \phi}\sum_{i=1}^N (r_i - C_j(x_i))\frac{ d_\phi(h_i|x_i)}{\hat{d}_{0}(h_i|x_i)} 
    + r_i\frac{ d_\phi(\bot|x)\hat{\pi}_\theta(a_i|x_i)}{\hat{d}_{0}(h_i|x_i)\hat{\pi}_{0}(a_i|x_i,h_i)}
\end{align}

\noindent where $\bot$ means algorithm is queried to make a decision, and $\hat{d}_0, \hat{\pi}_0$ are the estimated historical expert assignment and behavior model, respectively. We assume the human decision makers who produced the observational data are the same ones to whom decisions will be assigned in the future. It is possible, however, that new decision makers may enter the system while others may leave, and we leave it for future work. Here we also make the simplified assumption that assignment of human decision makers has no unobserved confounder and is the same as $x$. This assumption is frequently met; for instance, assignment of judges to cases is often randomized (in such cases $d_0 = 1/K$,  $K$ is the total number of human decision makers), and variables that influence non-random assignments, such as type of insurance in healthcare, are frequently contained in the data.

\section{Experiments}
In this section, we evaluate our proposed algorithms on both synthetic and real data that represent human decision making processes\footnote{Code and Appendix is available at \url{https://github.com/ruijiang81/hai-blbf}}. 
First, following the common practice for evaluating BLBF methods ~\cite{swaminathan2015counterfactual}, we convert multi-label supervised learning datasets to bandit feedback by creating a logging policy. More specifically, each instance $x$ has a label $y\in \{0,1\}^l$ where $l$ is the number of possible labels. After the logging policy selects an action $i$, a reward $y_i$ is revealed as a bandit feedback $(x,i,y_i)$, \ie,  for each data point, if the policy selects one of the correct labels of that data point, it gets a reward of 1 and 0 otherwise. By generating data this way, we have the full knowledge of counterfactual outcomes for evaluation. We use both synthetic and real human responses for evaluation. 

The data statistics of the five datasets used in our experiments are included in Appendix, and we discuss the detailed setup below. For all of the main experiments, the model is a three-layer neural network, the router model is another three-layer neural network. We later explore the effect of the number of hidden neurons for the model. We use Adam~\cite{kingma2014adam} with learning rate of 0.001 for optimization and both policy and router models are deterministic at testing time. We use truncated importance sampling estimator and set the truncation threshold as 10, and we select the baseline in~\cite{joachims2018deep} through a grid search over $[0,0.2,0.4,0.6,0.8]$ on the training data. The logging probability are estimated by an additional random forest model trained on observational data. We note that while there are alternatives to improve the logging probability estimation, this is beyond the scope of this paper. We train each method enough epochs until convergence.

\noindent \textbf{Algorithms}
We consider the following decision-making setups in our experiment. For all baselines without personalization, human experts are queried randomly.

\begin{itemize}
    \item Human: This baseline exclusively queries humans for decisions in the test set. 
    \item Algorithm Only (AO): This baseline trains the algorithm using Equation~\ref{eqn:ips}, and exclusively queries the algorithm for decisions in the test set. 
    \item Two Stage Collaboration (TS): The objective of the two stage baseline is shown in Equation~\ref{eqn:2s}. The policy for two stage is the same as AO; during test time, a routing model  decides whether to query the human or the algorithm for a decision. 
    \item Joint Collaboration (JC): The objective jointly optimizes both the target policy and routing algorithm. Here, the target policy is trained together with the routing model; during test time, a routing model  decides whether to query the human or the algorithm for a decision. 
    \item Joint Collaboration with Personalization (JCP): This algorithm corresponds to the objective in Equation~\ref{eqn:jointp}. When jointly learning the target policy and routing model, heterogeneous expertise are considered. At testing time, the routing model has the option to assign the new decision instance either to the algorithm or to one of the different human decision makers. 
\end{itemize}

\noindent \textbf{Simulated Data}
We show our experimental results for simulated human decision models on two multi-label datasets, Scene and TMC from  \href{https://www.csie.ntu.edu.tw/~cjlin/libsvmtools/datasets/multilabel.html}{LIBSVM repository}~\cite{elisseeff2002kernel,boutell2004learning}, which are used for semantic scene and text classification. Since we do not have the real human decisions on this dataset, we use two different human behavior models (HBM) to evaluate our methods. First, we fit  black box models on different subsets of 30\% of the data with full (not partial) ground truth labels and use it to generate simulated human decisions at both training and testing time. Note that this HBM can exhibit different decision accuracies for different instances. We use random forest (default implementation in scikit-learn package) as the black box model. The second HBM is motivated by labeling noise, where we assume each expert has an uniform decision noise across all instances, such that for any given instance, the expert will make the correct decision with a probability $\rho$ (yielding a reward of 1), and an incorrect decision with a probability $1-\rho$, resulting in zero reward. The human's decision is then drawn at  random from this distribution. Note that this is similar to classification noise, traditionally characterized by confusion matrix. For our experiments, we simulate three experts, setting $\rho$  as 0.6, 0.7, 0.8, for each expert respectively. Thus, experts exhibit differential decision-making accuracies and all have a much higher decision accuracy than random. The implementation details of both HBMs are included in the Appendix. Note that we assume a human decision maker behaves similarly across time and exhibits a good decision accuracy for the task, which is reasonable in settings where human experts make decisions, such as judges and physicians. Moreover, since human decision-makers do not interact with the algorithm in the proposed collaboration system, the decisions made by humans are likely to remain consistent before and after deployment of the system. In each experiment, we assume there are three experts present and the observational data is generated by a random assignment of these experts, which corresponds to settings such as the random assignment of judges to cases~\cite{ensign2018decision}. At test time, the routing algorithm queries certain expert or algorithm based on output probabilities randomly, then the decision-maker makes an action, instead of a uniform random assignment over all experts used in the collaboration objective. 

To create the observational data, given an expert's probability for each decision (determined by either the black-box HBM or the decision noise HBM), we sample the decision according to that probability. At testing time, the system chooses either to output the human's decision (using the black-box model to output an action) with probability $d_\phi$, or otherwise the trained policy is queried to offer a decision. For each decision, we observe a reward of 1 or 0 according to the ground-truth label in the original dataset, and we compare the total reward that each system achieves on the test set. In the main results, we set the human decision cost at $C(x) = C = 0.3$, and we later vary this cost in the ablation studies. Each experiment is run over ten repetitions and we report the average reward and standard error in Table~\ref{tab:simu}.

As shown in Table~\ref{tab:simu}, we find that when the models are jointly optimized, the human-machine team performs significantly better than the algorithm or human alone, whereas the two stage method does not always offer a significant improvement. 
Interestingly, we further observe that the human-machine team with personalized routing has a significantly better performance over both the baselines of human or algorithm alone, which confirms our intuition and the potential of human-AI collaboration with diverse human experts. 

\begin{table}[h]
    \centering
    \resizebox{\linewidth}{!}{
\begin{tabular}{lrrrrr}\hline
     &  Human & AO & TS& JC & JCP\\ \hline
    Scene (Model) & 341.3$\pm$7.9 & 376.3 $\pm$9.3& 379.2$\pm$9.2&423.3$\pm$5.2&425.4$\pm$4.2 \\
    Scene (Noise) & 294.8$\pm$12.0& 358.1$\pm$7.8&364.7$\pm$7.7&391.9$\pm$8.4&408.3$\pm$7.5 \\
    TMC (Model) & 4919.5$\pm$16.2 & 5543.7$\pm$109.2& 5642.9$\pm$92.2&5736.1$\pm$87.5&5787.8$\pm$92.0\\
    TMC (Noise) & 3435.1$\pm$28.3 & 4438.1$\pm$131.5&4361.5$\pm$113.6&4513.6$\pm$87.9&4935.1$\pm$94.3\\
    \hline
\end{tabular}}
    \caption{Reward on simulated dataset for different HBMs.}
    \label{tab:simu}
\end{table}

\noindent \textbf{Real Data}
We evaluate our approach with real human decisions, using the data used in ~\cite{li2018multi} for Multi-Label Learning (MLC) and a sentiment analysis dataset (Focus)~\cite{rzhetsky2009get} from crowd workers. The data statistics are shown in Section B in Appendix. MLC has annotations from 18 workers, each annotating at least 70 images. On average, each worker annotated 267 $\pm$ 201 images, and each image was annotated by 6.9 $\pm$ 2.3 workers. Because not all workers worked on the same instances, we cannot query every worker for our personalization experiment; hence, for MLC, we view all 18 workers as a group of workers. Specifically, when querying from a human, a  worker is sampled at random from the workers who labeled this instance to offer a decision. The action (decision) made by this worker is sampled from the worker's chosen label(s). For example, if the drawn worker labeled an instance with true label \{tree, beach\} as \{beach,sea\}, a random decision of \textit{beach} might be sampled as an action, yielding a reward of 1.\footnote{We note that 5 instances in MLC had no annotation, and we remove them from the dataset.} We set the test set ratio as 15\% and randomly sampled five training-testing splits to validate our proposed algorithms. We train each human-machine system on the generated observational data. A reward of 1 is produced when the decision is correct, and the reward is 0 otherwise. Significance test results are included in Appendix. 

Focus is a binary classification dataset where each instance is labeled independently by five crowd workers, allowing us to evaluate our personalization objective using real human responses. We follow the same experiment setup above to convert it into a decision-making dataset, and set the test set size as 30\%. The average decision accuracy of five workers are shown in Table~\ref{tab:focus}, and is calculated as the percentage of accurate decisions for all instances. Most workers demonstrate good decision accuracies in the dataset, but the Worker 5 has a relatively lower decision accuracy compared to others, which aligns with our motivation for designing the personalized routing objective. For the same cost of 0.3, we notice that human experts have much lower reward than the algorithm, and the human-machine team learns to only output algorithm decisions, thus we set cost to a range of different values from 0 to 0.5 in order to check how different costs will affect each method. A cost of 0 can correspond to a high stake domain where accuracy is the ultimate goal regardless of any human expert cost. 

\begin{table}[h]
    \centering
    \scalebox{0.8}{
    \begin{tabular}{lrrrrr}
    \hline
     Worder ID &  1 & 2 & 3 & 4 & 5 \\\hline
     Decision Accuracy & 0.82 & 0.93 & 0.89 & 0.90 & 0.64 \\\hline
    \end{tabular}}
    \caption{Worker decision accuracy for the Focus dataset.}
    \label{tab:focus}
\end{table}

The reward generated by each baseline on the test set is summarized in Table~\ref{tab:real}. Similar to the results on the simulated data, we find the hybrid team outperforms the baseline where an algorithm works alone, and the joint optimization further improves over the two stage approach for a lower cost. Interestingly, we also find that when the cost is higher, the human-machine team will learn to only output algorithm decisions and achieve the same performance as AO baseline. These results further confirm the benefit of our proposed approach with real datasets annotated by human workers. We also find that in the Focus dataset, JCP allocates all instances only to Workers 1-4, and none to Worker 5 across repetitions, which shows the benefit of personalized routing. 


\begin{table}[h]
    \centering
    \resizebox{1\linewidth}{!}{
    \begin{tabular}{lrrrrr}\hline
         Data (cost) &  Human & AO & TS& JC& JCP\\ \hline
        MLC (0.3) & 53.8$\pm$3.6&66.5$\pm$1.1&77.1$\pm$0.7&79.2$\pm$1.0& ---\\ 
        Focus (0) & 250.3 $\pm$4.2& 231.2$\pm$2.0&231.5$\pm$2.2&250.3$\pm$1.3&270.4$\pm$2.0 \\ 
        Focus (0.05) & 235.3$\pm$4.2&231.2$\pm$2.0&231.3$\pm$2.0&237.5$\pm$1.8&257.3$\pm$1.9\\ 
        Focus (0.1) & 220.3$\pm$4.2 & 231.2$\pm$2.0 & 231.0$\pm$2.0& 239.7$\pm$1.4&245.2$\pm$1.5 \\
        Focus (0.3) &160.3$\pm$4.2 & 231.2$\pm$2.0&231.2$\pm$2.0&230.33$\pm$1.8&229.8$\pm$1.4\\
        Focus (0.5) & 100.3$\pm$4.2&231.2$\pm$2.0&231.2$\pm$2.0&231.2$\pm$1.9&230.3 $\pm$1.5 \\ 
        \hline
        \end{tabular}}
    \caption{Reward on real datasets with different expert costs.}
    \label{tab:real}
\end{table}

\noindent\textbf{Hybrid System Improvement}
In this section, we conduct ablation studies to better understand the limitations and how model selection affects the human-machine team performance. We use our real dataset MLC from above to better understand when human-machine collaboration will achieve a greater improvement. Intuitively, if the algorithm alone can achieve perfect performance, a human-machine team will not improve performance with respect to the optimized objective. 

We fix the routing model's architecture as before and set the policy model to different number of neurons to represent different model capacities, since by universal approximation theorem, with a sufficiently large non-linear layer, we can approximate any continuous function $f$~\cite{bengio2017deep}. We run each experiment for 10 runs, also the number of neurons is in a small range to avoid potential overfitting since the dataset is not difficult to learn. The average rewards across algorithms are reported in Table~\ref{tab:ablationreal}. We also add in relu activation after linear layer as an alternative to increase model capacity. When the model has limited capacity (2 units), we observe a greater benefit of using human-machine team. When the model has a stronger capacity, we observe a stronger performance in the AO baseline, and the joint optimization and hybrid team's improvement drops. This finding confirms our intuition that the human-machine collaboration helps to a lesser degree when the model has a strong capacity and no overfitting happens. 

\begin{table}[h]
    \centering
    \scalebox{0.75}{
    \begin{tabular}{lrrr}\hline
        \# Hidden Units & AO & TS& JC\\ \hline
        $2$ & 66.5$\pm$1.1&77.1$\pm$0.7&79.2$\pm$1.0\\ 
        $2$ (w/ \texttt{ReLU}) & 66.9$\pm$1.2&75.9$\pm$1.6&77.7$\pm$1.3 \\
        $8$  & 77.7 $\pm$1.0&86.3$\pm$0.7&86.4$\pm$0.9\\ \hline
    \end{tabular}}
    \caption{Rewards on different model capacity on MLC.}
    \label{tab:ablationreal}
\end{table}

\section{Conclusion and Future Work}
In this paper, 
we propose a novel collaborative objective for human-machine collaboration and a personalized routing objective when there are multiple decision-makers present. Our empirical evaluation validates the effectiveness and benefit of our methods on both simulated and real human decisions. 

There are many potential extensions to our proposed human-machine collaboration system. First, experts might perform better than algorithms when the underlying data distribution has shifted; the algorithm has no chance to learn from it, while humans may generalize better for these cases. Second, humans may have access to some additional features that are not available to the algorithm; humans will perform better than the algorithm on such cases. How to model the logging policy under such settings becomes challenging. Moreover, here we assume that the historical data is generated by human decision makers, and each instance is only decided by one human decision maker. However, in practice, many decision-making systems involve group deliberation, such as panel discussions where multiple decision-makers make the final decision. Hybrid human-AI decision making may also become increasingly common. Making inferences from historical data in such settings is more challenging and we leave them for future work. 

\section*{Acknowledgements}
We thank the many talented Amazon Mechanical Turk workers who contributed to our study and made this work possible. We also thank our reviewers for their valuable feedback. This research was supported in part by Wipro, a Google AI Award for Inclusion Research, the Micron Foundation, and Good Systems\footnote{\url{https://goodsystems.utexas.edu}}, a UT Austin Grand Challenge to develop responsible AI technologies. The statements made herein are solely the opinions of the authors.

\bibliography{example_paper}
\bibliographystyle{named}

\end{document}